\documentclass[aps,prl,twocolumn,superscriptaddress,showpacs,nobibnotes]{revtex4}

\usepackage{graphicx}
\usepackage{bm}
\usepackage{subfigure}
\usepackage{latexsym}
\usepackage{placeins}
\usepackage{amstext,amssymb,amsfonts}
\usepackage{epsfig}
\usepackage{color}
\usepackage{titlesec}
\usepackage{amsmath}
\usepackage{float}
\usepackage{hyperref}

\usepackage[dvipsnames]{xcolor}

\begin{document}

\title{Design of endurable networks in the presence of aging}

\author{Yuansheng Lin}
\thanks{These two authors contributed equally to this work.}
\affiliation{School of Reliability and Systems Engineering, Beihang University, Beijing, China, 100191}
\affiliation{Science and Technology on Reliability and Environmental Engineering Laboratory, Beijing, China, 100191}
\affiliation{Department of Computer Science, University of California, Davis, California, USA, 95616}

\author{Amikam Patron}
\thanks{These two authors contributed equally to this work.}
\affiliation{Department of Physics, Bar-Ilan University, Ramat-Gan 5290002, Israel
}

\author{Shu Guo}
\affiliation{School of Reliability and Systems Engineering, Beihang University, Beijing, China, 100191}
\affiliation{Science and Technology on Reliability and Environmental Engineering Laboratory, Beijing, China, 100191}

\author{Rui Kang}
\affiliation{School of Reliability and Systems Engineering, Beihang University, Beijing, China, 100191}
\affiliation{Science and Technology on Reliability and Environmental Engineering Laboratory, Beijing, China, 100191}

\author{Daqing Li}
\thanks{Corresponding author: Daqing Li, daqingl@buaa.edu.cn}
\affiliation{School of Reliability and Systems Engineering, Beihang University, Beijing, China, 100191}
\affiliation{Science and Technology on Reliability and Environmental Engineering Laboratory, Beijing, China, 100191}

\author{Shlomo Havlin}
\affiliation{Department of Physics, Bar-Ilan University, Ramat-Gan 5290002, Israel
}

\author{Reuven Cohen}
\affiliation{Department of Mathematics, Bar-Ilan University, Ramat-Gan 5290002, Israel}

\date{\today}
\begin{abstract}
Networks are designed to satisfy given objectives under specific requirements. While the static connectivity of networks is normally analyzed and corresponding design principles for static robustness are proposed, the challenge still remains of how to design endurable networks that maintain the required level of connectivity during its whole lifespan, against component aging. We introduce network endurance as a new concept to evaluate network’s overall performance during its whole lifespan, considering both network connectivity and network duration. We develop a framework for designing an endurable network by allocating the expected lifetimes of its components, given a limited budget. Based on percolation theory and simulation, we find that the maximal network endurance can be achieved with a quantitative balance between network duration and connectivity. For different endurance requirements, we find that the optimal design can be separated into two categories: strong dependence of lifetime on node's degree leads to larger network lifetime, while weak dependence generates stronger network connectivity. Our findings could help network design, by providing a quantitative prediction of network endurance based on network topology.
\end{abstract}
\maketitle

Robustness \cite{boccaletti-complex-2006,Dorogovtsev-RevModPhys-2008,strogatz-nature-2001,albert-RevModPhys-2002,albert-nature-2000,cohen-prl-2000,cohen-prl-2001,zio-RelEng-2016} is a prerequisite condition for system functionality in various types of network system designs, including critical infrastructures (e.g. power grids \cite{buldyrev-nature-2010,daqing-scirep-2014l}, communication networks \cite{Dodds-proc-nat-acad-sci-2003} and transportation networks \cite{li-proc-nat-acad-sci-2015,CHOWDHURY-phys-rep-2000}) and natural systems (e.g. biological systems \cite{kitano-nature-2004,kitano-science-2002}, ecological systems \cite{folke-glob-envir-chng-2006,anderies-ecology-society-2004} and social networks \cite{newman-nat-acad-sci-2002,eubank-nature-2004}). Robustness enables a system to perform fully or at least at an acceptable minimum function after a failure of a portion of its components, e.g. due to internal faults and/or external hazards. On the other hand, the structure of a system evolves during its life and its components fail due to aging, an effect that has been rarely considered. Indeed, aging processes occur in engineering systems \cite{li-trans-power-sys-2002,KOLOWROCKI-rel-eng-sys-2008}, biological systems \cite{finkel-nature-2000,Harman-proc-nat-acad-sci-1981}, and even online social systems \cite{viswanath-2009,Leskovec-2008}, where users may finally quit an online community after an active period. 

Static robustness of a network can be defined as its ability to withstand losses of nodes or links under random failures or targeted attacks \cite{albert-nature-2000,cohen-prl-2000,cohen-prl-2001,berezin-sci-rep-2015}. Based on percolation theory, network static robustness can be characterized by a percolation critical threshold, which is the critical fraction of failed network elements at which the system collapses. It has been shown \cite{albert-nature-2000,cohen-prl-2000,cohen-prl-2001} that scale-free networks are usually more robust than Erd\H{o}s-R\'{e}nyi (ER) networks with respect to random failures, but they are more fragile to targeted attacks. Efforts \cite{paul-europ-phys-journal-2004,shargel-prl-2003, tanizawa-pre-2005} have been made to find optimal designs of network structures that are robust to both random failures and targeted attacks. However, the optimization of robustness by connectivity design is not sufficient because the connectivity of real networks is not static and robustness changes also due to aging process of the components \cite{Braunewell-pre-2008,Klemm-proc-nat-acad-sci-2005,kauffman-physicaD-1986,Aldana-proc-nat-acad-sci-2003, Schneider-proc-nat-acad-sci-2011}. For example, cellular networks decline and may collapse due to the aging of several cells every minute and online social networks are suffering from the daily loss of users.

Whereas previous studies mostly focus on static network robustness at a given snapshot of its life \cite{albert-nature-2000,cohen-prl-2000,cohen-prl-2001,berezin-sci-rep-2015,Schneider-proc-nat-acad-sci-2011}, the effect of network evolution due to aging of components has rarely been addressed. In this paper, network endurance is proposed as a concept to evaluate the network’s overall performance during its whole lifespan, thus considering both network connectivity and network lifetime. Then, the following question becomes fundamental: given a constrained budget, how to optimally allocate resources of components' lifetimes in order to design an endurable network for its whole lifespan?

Network performance in the whole lifespan depends on the health of the system components, and also the connection between these healthy components. We develop an objective function $W$ (Eq. (\ref{eq:W}) below) for the optimization of network endurance during its whole lifespan. In our case, we assume to have only the information on the degrees of the network components in the design stage. For example, for Peer-to-Peer file sharing network (P2P), it is hard to know the whole topology \cite{saroiu-2002}, but it would be easier for a peer to have the information about its neighbors based on the communication mechanisms. Ref. \cite{Leonard-perf-eval-rev-2005} finds that node’s residual lifetime correlates with the number of node’s neighbors in a P2P systems. Thus, in our model, each node in the network is allocated with a lifetime expectancy, depending on its number of connected direct links. We, then, show how to design an endurable network, using both simulation and theoretical analysis. For this, we consider two exponents: $\alpha$, which characterizes the power-law relation between node expected lifetime and degree, and $\beta$, which measures the user’s requirement for network endurance.

\begin{figure}[t]
\begin{center}
  \includegraphics[width=1\columnwidth]{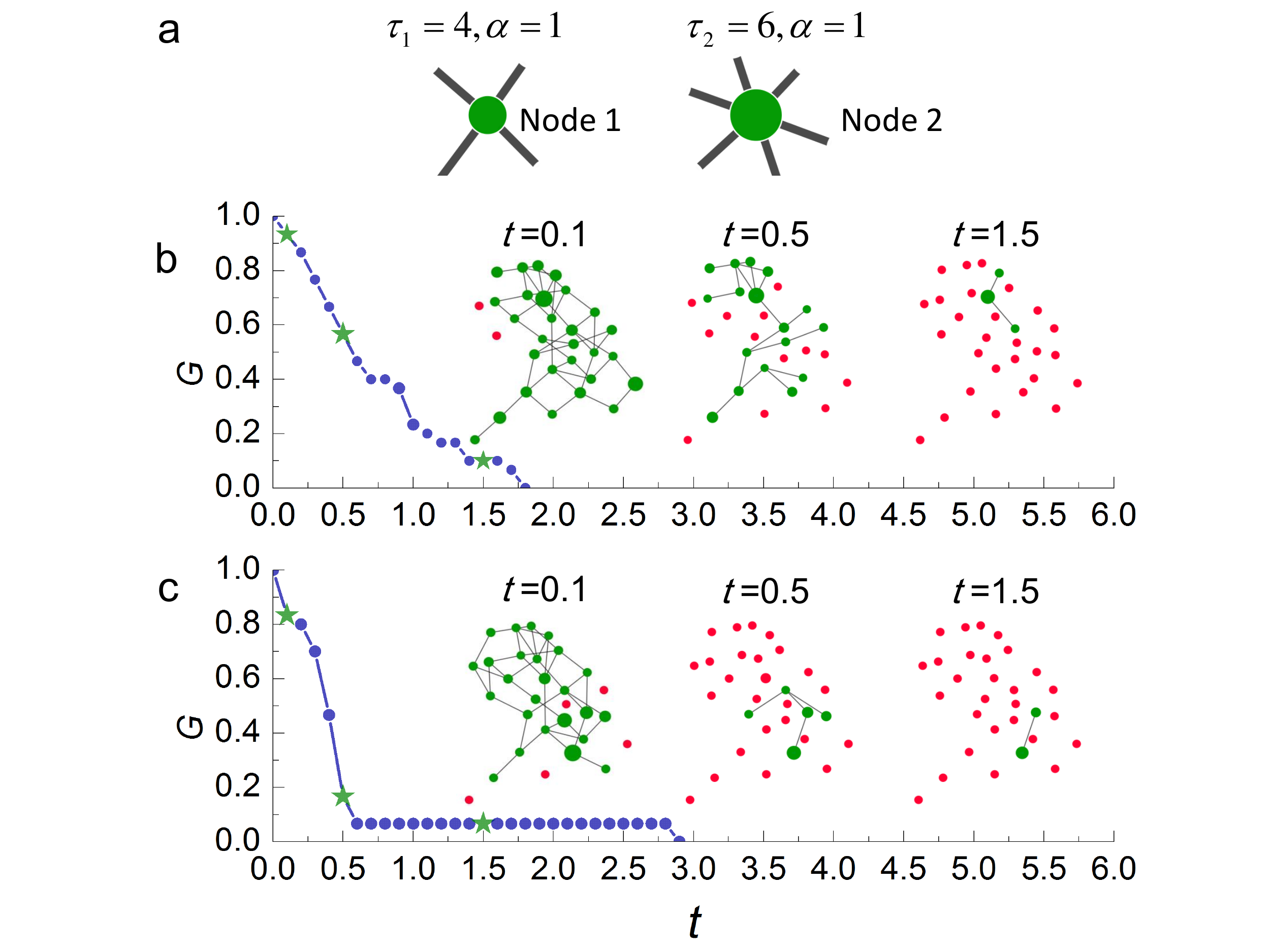}
\end{center}
\caption{Evolution of a network (giant component, $G$) with aging time $t$. (a) Node’s characteristic lifetime is assumed to depend on their degree as $\tau_i=k_i^\alpha$ ; (b) Time evolution of a network with large $G$ but short lifetime: the figure demonstrates the case of $\alpha=0.5$. In the figure, the size of a node (green circle) is proportional to its remaining lifetime. The shown scale-free (SF) network contains 30 nodes, and has a power-law exponent $\gamma=2.5$ and average degree $4$; (c) The same SF network but with $\alpha=2.5$, in this case, initially the network has a small $G$ but relatively longer lifetime compared to (b).}
\label{fig:introduction}
\end{figure}

As demonstrated in Fig. \ref{fig:introduction}a, for a given node $i$ with degree $k_i$, we allocate a characteristic lifetime, according to the relation $\tau_i=k_i^\alpha$. In Fig. \ref{fig:introduction}b and Fig. \ref{fig:introduction}c, we represent the aging process by the evolving snapshots of the network at successive times. In Fig. \ref{fig:introduction}b, the network has in the beginning a large connected network component but a short lifetime, while in Fig. \ref{fig:introduction}c the network is comparatively smaller than that in Fig. \ref{fig:introduction}b but has a longer lifespan.

In practice, the lifetimes of the components, e.g. electronic components \cite{porschan-technometrics-1963,pecht-2008} are not accurately determined and in reliability analysis it is common to assume a lifetime exponential distribution with mean value $\tau$ \cite{zio-2007}. Another realistic lifetime distribution is the Weibull distribution, which is realistic for many mechanical components \cite{weibull-jrnl-appl-mech-1951}. For the exponential lifetime distribution, the survival probability of a node is $e^{-t/\tau}$ at time $t$. In our model, as mentioned earlier, we assume that the mean value of node lifetime depends only on its degree, according to a power-law relation $\tau=k^\alpha$, where $\alpha$ is the exponent to be optimized for network endurance. Fig. \ref{fig:introduction}a illustrates the lifetime allocation of nodes in networks. Determining the optimal $\alpha$ will tell us how to distribute lifetimes (cost) between the nodes of different degrees.

Depending on the application, the objective of network endurance is to have a large $G$ as long as possible. However, the large size of $G$ and its long lifetime are in competition due to the limited total cost. To describe these competing processes, we define the endurance function $W$,
\begin{equation}
\label{eq:W}
W=\int G\left(t\right)^\beta dt,
\end{equation}
which evaluates the network capacity of maintaining the connectivity function through the whole lifespan. The value of $W$ integrates the overall performance of the network in terms of endurance. The exponent $\beta$ controls the importance of the size based on the design requirement for network endurance. For example, $\beta=0$ corresponds to the case where a maximal duration of the network is required, regardless of the connectivity performance during this lifespan (i.e. having a giant component). When $\beta=1$, the size of the giant component during the network lifespan is also taken into account. As $\beta$ increases, more weight is given to the network giant component size as design requirement, compared to the network lifetime. When $\beta$ approaches large values ($\beta>>1$), the design requirement for network endurance is focused on the size of the giant component. Our aim is then to find the optimal $\alpha$ (for a given $\beta$) that maximizes $W$.

We define $f(t|k)$ to be the conditional density probability of a node to have a lifetime $t$, given its degree is $k$. The expected lifetime of a node of degree $k$, that is also the expectation of $f(t|k)$, is $\tau(k)$, which assumed to be proportional to $k^\alpha$. We also assume that the total budget of lifetime of all components is equals to the network size $N$. Thus, $\tau(k)=N\frac{k^\alpha}{\sum_{k=0}^\infty k^\alpha p(k)N}=\frac{k^\alpha}{\sum_{k=0}^\infty k^\alpha p(k)}$. We define $q(k,t)$ as the probability that a randomly chosen node that has a degree k survives at time $t$. We can calculate it as follows (take the exponential distribution for example)
\begin{equation}
q(k,t)=p(k)\int_t^\infty f(t^\prime|k)dt^\prime=p(k)\int_t^\infty \frac{1}{\tau(k)}e^{-t^\prime/\tau(k)}dt^\prime.
\end{equation}

Next, we add a time parameter, $t$, to the percolation generating functions \cite{callaway-prl-2000}, and the generating function of a node that survives at time $t$ is
\begin{equation}
F_0(x,t)=\sum_{k=1}^\infty q(k,t)x^k.
\label{eq:F0}
\end{equation}
The probability that a randomly chosen edge leads to a node, survives at time $t$, is $\frac{(k+1)q(k+1,t)}{\langle k \rangle}$, where $\langle k \rangle$ is the network mean degree. And the corresponding generating function is 
\begin{equation}
F_1(x,t)=\sum_{k=1}^\infty \frac{kq(k,t)}{\langle k \rangle}x^{k-1}.
\end{equation}
Hence, the generating functions for the component size that all its nodes survive at time $t$ is,
\begin{align}
H_1(x,t)=1-F_1(1,t)+xF_1\left[H_1\left(x,t\right),t\right].\nonumber\\
H_0(x,t)=1-F_0(1,t)+xF_0\left[H_1\left(x,t\right),t\right].
\label{eq:H01}
\end{align}
The size of the giant component at time $t$ is $G(t)=1-H_0(1,t)$, since $H_0(1,t)$ contains only finite size components that survive in time $t$. Thus, using Eqs. ($\ref{eq:H01}$) and ($\ref{eq:F0}$) we get
\begin{equation}
G(t)=F_0(1,t)-F_0\left[H_1\left(1,t\right),t\right]=\sum_{k=0}^\infty q(k,t)\left(1-u(t)^{k}\right),
\label{eq:G_t}
\end{equation}
where $u(t)\equiv H_1\left(1,t\right)$. 
From Eqs. (\ref{eq:G_t}) and (\ref{eq:H01}) we calculate the endurance $W$ (Eq.(\ref{eq:W})), by solving numerically the following equations:
\begin{align}
W&=\int_{t=0}^{\infty}\left[\sum_{k=0}^{\infty} q(k,t)\left(1-u(t)^{k}\right)\right]^\beta dt\nonumber\\
u(t)&=1-\sum_{k=1}^{\infty} \frac{kq(k,t)}{\langle k \rangle}\left[1-u(t)^{k-1}\right].
\label{eq:W_genfun}
\end{align}

\begin{figure}[t]
\begin{center}
	  \includegraphics[width=\linewidth]{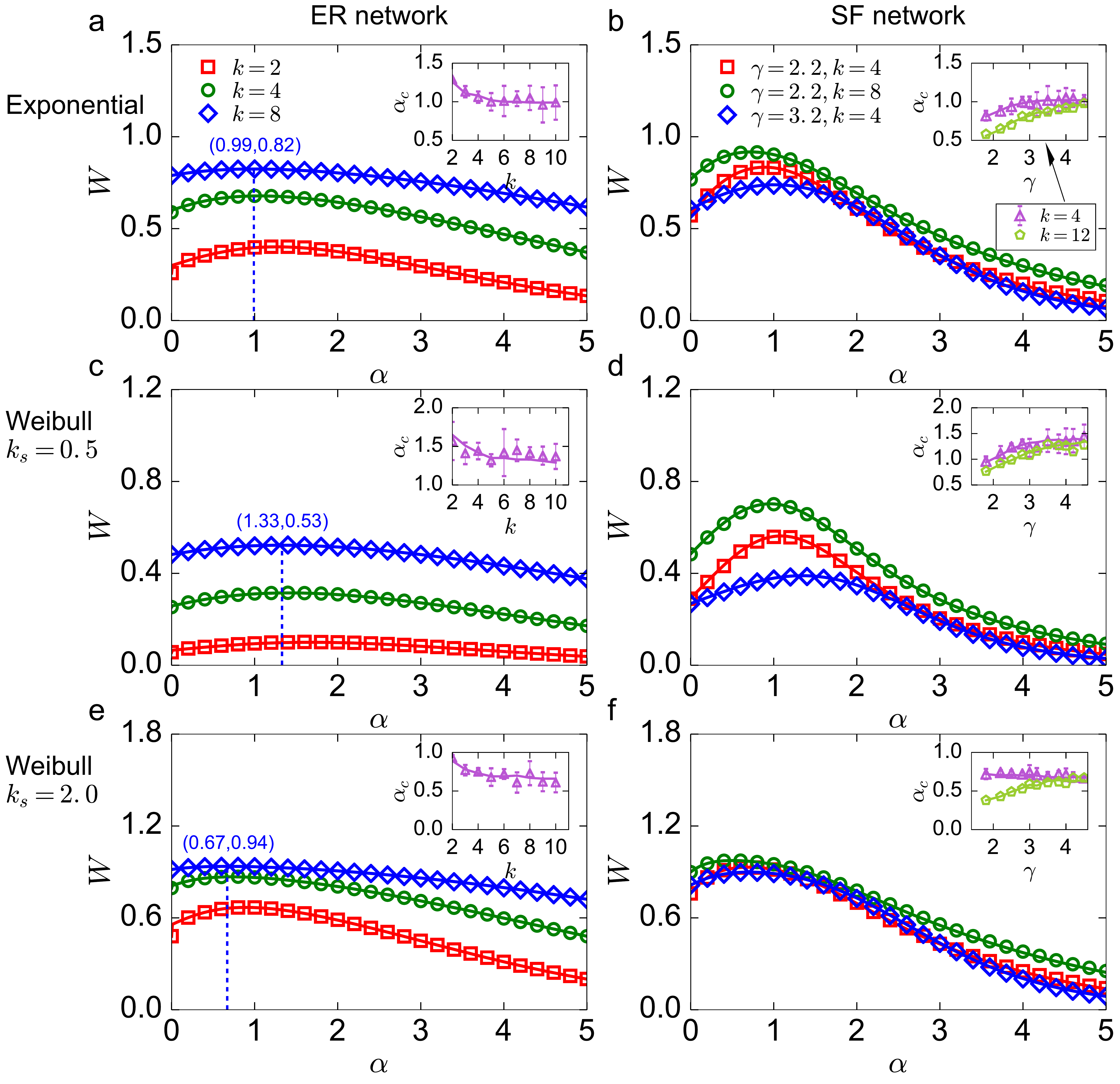}
\end{center}
\caption{The dependence of endurance $W$ on $\alpha$ for different networks and component lifetime distributions ($\beta=1$). (a,b) The case of exponential lifetime distribution for component lifetime: (a) Results for $W$ of ER networks with different average degree $k$. Open symbols represent the simulation results in networks with $N=10^4$ nodes and averaged over $500$ realizations, while solid lines are obtained from the theoretical predictions, Eqs (\ref{eq:W_genfun}). Both simulations and theoretical solutions were implemented by a summation of the giant component size over time in time steps of $0.01$ units (The integral in Eq. (\ref{eq:W_genfun}) was replaced by a summation). Inset: the relationship between $\alpha_c$ and $k$ is shown. The error bars are the standard deviations for $\alpha_c$, calculated from $10$ single realizations. (b) Results for SF networks with different power-law exponents but same average degree ($k=4$ or $k=8$). The network contains $N=10^4$ nodes. Inset: the relationship between $\alpha_c$ and $\gamma$ for scale-free networks. (c,d) Weibull lifetime distribution case with shape parameter, $k_s=0.5$, for ER and SF networks, which generates a broader distribution compared to the exponential distribution. (e,f) Weibull lifetime distribution case with shape parameter $k_s=2.0$ for ER and SF networks: the lifetime is narrower than for exponential distribution. The reason for $\alpha_c>1$ for small $k$ (in a, c and e) is due to the fact that some isolated clusters exist, where some long lifetime allocation investment is wasted. Since high degree nodes are less probable to be in small clusters, the network functionality will gain more when allocating longer lifetimes to them.}
\label{fig:W_vs_alpha}
\end{figure}

We begin by analyzing the case where the lifetime of components following an exponential distribution. We consider firstly the case of $\beta=1$, and study how the exponent $\alpha$ affects the performance of network endurance $W$. In Fig. \ref{fig:W_vs_alpha}a, we show the endurance $W$ for the ER network as a function of $\alpha$. We can see that for an ER network with a given average degree, $W$ increases gradually as $\alpha$ increases and reaches the maximum at $\alpha=\alpha_c$. For example, for an ER network with $k=4$, $\alpha_c\cong 1$. After the maximum, investing more resources on nodes with large degree will lead to the early failure of nodes with small degree, which will decrease the network endurance. Furthermore, Fig. \ref{fig:W_vs_alpha}a shows that ER networks with larger average degrees have larger giant components, leading to larger network endurance. Interestingly, as seen in the inset of Fig. \ref{fig:W_vs_alpha}a, for $\beta=1$, the optimal $\alpha$ saturates at a constant value around $1$, for average degree above $4$. This stable design configuration is reached due to the competition between high-degree nodes and low-degree nodes in the network design. On one hand, high-degree nodes are more critical than low-degree nodes for the network integration; on the other hand, low-degree nodes may play a role of weak ties connecting different components to form a giant component \cite{morone-nature-2015}.

A SF network has a more heterogeneous structure than an ER network: some nodes could have very large degree (hubs), but most have only a few connected neighbors. To optimize SF network endurance, the balance between hubs and lower degree nodes becomes more sensitive. As shown in Fig. \ref{fig:W_vs_alpha}b, SF networks display a sharper maximum for $W$, compared to ER networks. We find that for SF networks with small power-law exponent $\gamma$, the optimal $\alpha$ is smaller than $1.0$ found in ER networks (See Inset of Fig. \ref{fig:W_vs_alpha}b). This is because for similar $\alpha$ values, hubs in SF networks are usually much larger and will receive much longer lifetime than in ER networks, while low-degree nodes with shorter lifetime will fail in the early stages of network evolution. Meanwhile, $\alpha_c$  increases within a narrow range between $0.5$ and $1.0$ in SF networks with increasing power-law exponent $\gamma$, and finally approaches $1$ (as for ER networks) when $\gamma$ is close to $4.0$. We also find that $\alpha_c$ increases with decreasing average degree.

\begin{figure}[t]
\begin{center}
	\includegraphics[width=\linewidth]{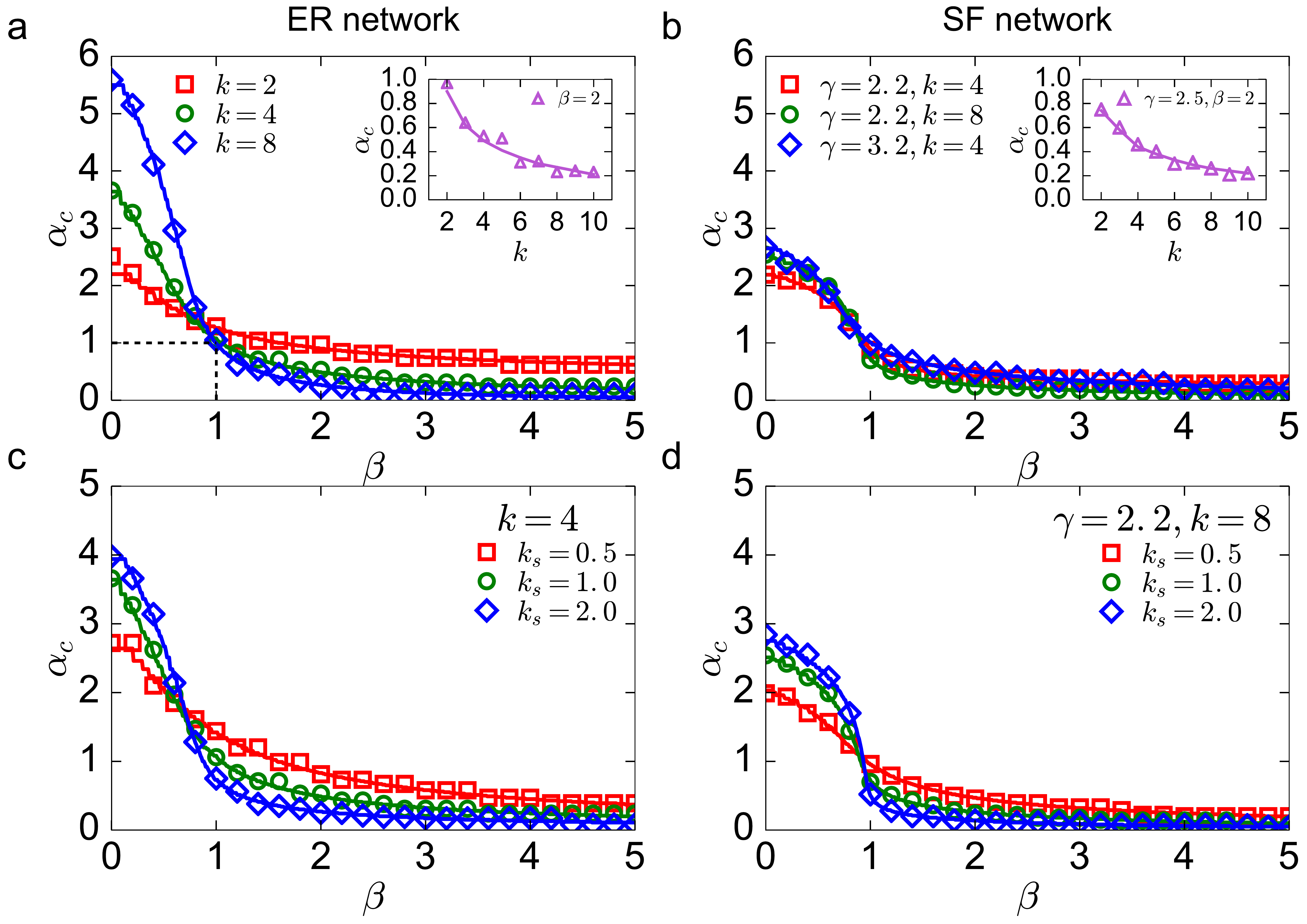}
\end{center}
\caption{The dependence of $\alpha_c$ on $\beta$. For the case of exponential lifetime distribution: (a) Simulation (symbols) and theoretical (lines) results for ER networks with different values of average degree $k$ and the network size is $10^4$; Inset: for $\beta=2$, the relationship between $\alpha_c$ and $k$ is shown. (b) Simulation (symbols) and theoretical (lines) results for different SF networks. The SF network contains $10^4$ nodes. Inset: for $\beta=2$, the relationship between $\alpha_c$ and $k$ for SF networks with $\gamma=2.5$ is shown. We compare the effects of different component lifetime distributions on the values of $\alpha_c$: (c) ER network, $k=4$, and (d) SF network, $\gamma=2.2, k=8$.}
\label{fig:alphac_vs_beta}
\end{figure}

\begin{figure}[t]
\begin{center}
	\includegraphics[width=\linewidth]{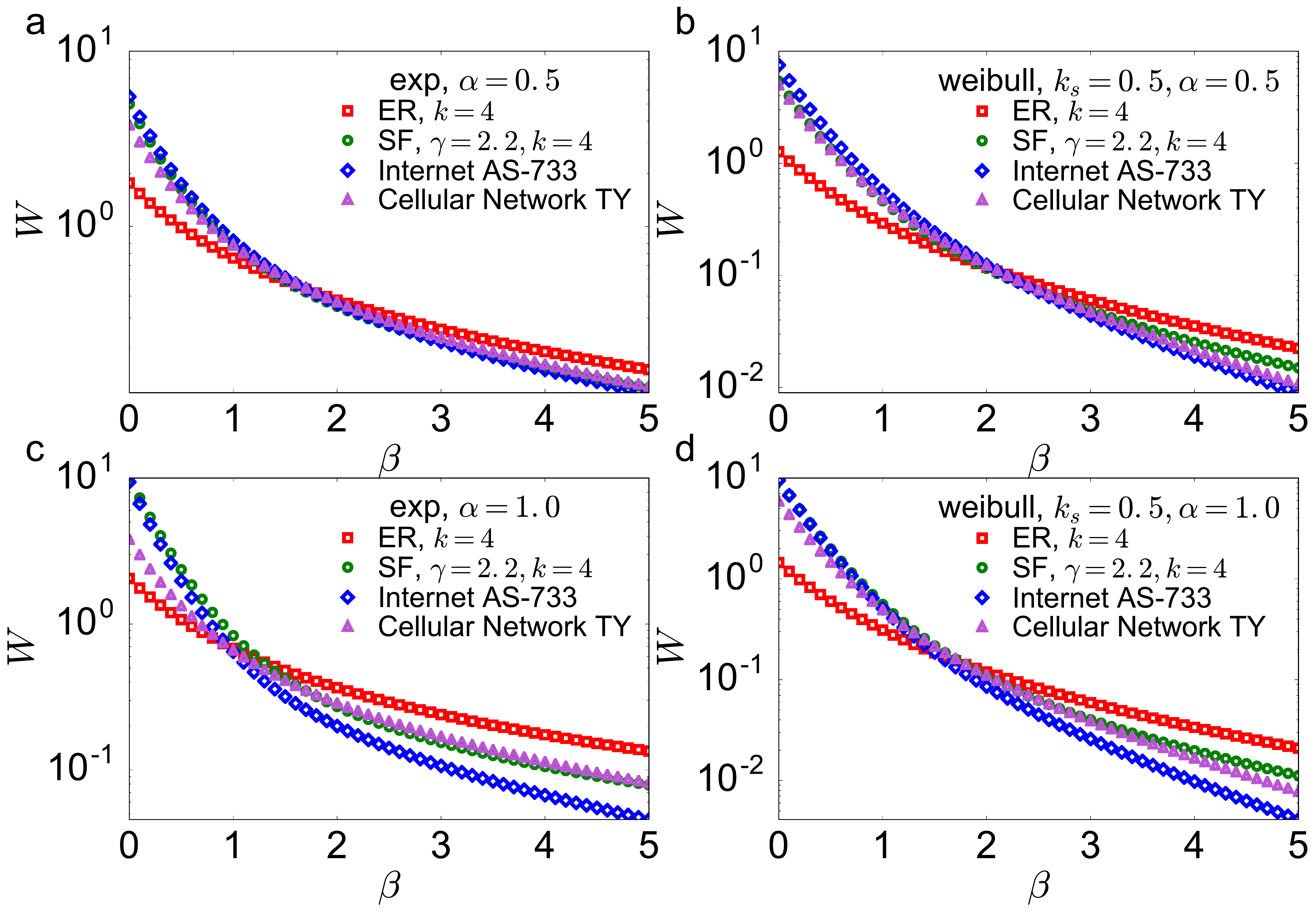}
\end{center}
\caption{The dependence of endurance $W$ on $\beta$ for different networks and component lifetime distributions. We calculate endurance for two network models (ER network and SF network), and two real networks (Internet AS-733 and Cellular Network (Salmonella typhi)). In (a) and (c) ($\alpha=0.5$ and $\alpha=1.0$, respectively), the lifetime of network components follows the exponential distribution. In (b) and (d) ($\alpha=0.5$ and $\alpha=1.0$, respectively), the lifetime of network components follows the Weibull distribution ($k_s=0.5$).}
\label{fig:4}
\end{figure}

For Weibull lifetime distributions, at time $t$ the survival probability for a component follows $e^{-\left(t/\lambda\right)^{k_s}}$, where $\lambda$ is the scale parameter, $k_s$ is the shape parameter, and the lifetime expectancy is $\lambda\Gamma\left(1+\frac{1}{k_s}\right)$ (in our case it is proportional to $k^\alpha$). The broadness of the Weibull distribution is controlled by the shape parameter $k_s$. From Fig. \ref{fig:W_vs_alpha}c, for $k_s=0.5$, where the lifetime distribution is relative broad, we find that for $\beta=1$, ER networks achieve optimal endurance for $\alpha$ larger than $1.0$. For SF networks, in Fig. \ref{fig:W_vs_alpha}d, we also see that their endurance is also optimized at larger values of $\alpha_c$. In these cases, in order to achieve the optimal resources allocation for network endurance, nodes with high degrees deserve more allocated resources. In Fig. \ref{fig:W_vs_alpha}d, we also find that the peak of network endurance for SF networks is sharper than for ER networks (Fig. \ref{fig:W_vs_alpha}c).

For $k_s=2$, where the lifetime distribution is narrower, we find that the optimal endurance for ER and SF networks are obtained at smaller values of $\alpha_c$, as shown in Fig. \ref{fig:W_vs_alpha}e and Fig. \ref{fig:W_vs_alpha}f, suggesting that networks’ resources should be shared more equally to optimize endurance.

Next, we study how the users' requirements, represented by $\beta$,  influences the optimal design. We find in Fig. \ref{fig:alphac_vs_beta}a and Fig. \ref{fig:alphac_vs_beta}b that the optimal $\alpha_c$ decreases with increasing $\beta$ for both ER and SF networks. When we put more weight on the network connectivity by increasing $\beta$, resources should be more uniformly invested among nodes of different degrees, which is represented by lower $\alpha_c$. Indeed, when we are interested in only the network connectivity at large values of $\beta$, we obtain very small values for $\alpha_c$ (see Fig. \ref{fig:alphac_vs_beta}): approaching the design result for a static network, neglecting the effect of aging. Meanwhile, when the network lifetime is also considered important with decreasing $\beta$, large degree nodes need to function in order to bridge different components in the giant component. Therefore, $\alpha_c$ increases continuously with decreasing $\beta$. Interestingly, we find that in both ER and SF networks the optimal $\alpha_c$ is close to $1$ for $\beta$ close to $1.0$ for different combinations of network parameters. 

Moreover, we find that different lifetime distributions of components have significant effects on the value of $\alpha_c$ for the same network (Fig. \ref{fig:alphac_vs_beta}c and Fig. \ref{fig:alphac_vs_beta}d): when $\beta$ is small, narrow lifetime distribution (e.g. $k_s=2.0$) leads to larger value of $\alpha_c$ compared with broad lifetime distribution (e.g. $k_s=0.5$). However, when $\beta$ increases to a certain value, the situation is reversed, the value of $\alpha_c$ becomes smaller for narrow lifetime distribution but higher for broad lifetime distribution.

In Fig. \ref{fig:4}, we also present how network endurance changes with users' requirements. We can find that both for network models and real networks, when $\beta$ increases, their network endurance will decrease. Moreover, Figure 4 shows that network with more heterogeneous degree distribution (for example, Internet AS-733 and SF network), will have larger network endurance when users focus more on network lifetime. But when network connectivity is preferred, network with homogeneous degree distribution (e.g. ER network) will gain more network endurance. 


To summarize, in this work we proposed a new realistic concept for the robustness of a random network, by considering the functionality of the network during its whole lifespan, instead of the traditional approaches valid only for networks at a given snapshot. Here the robustness is calculated in the presence of components \textit{aging} by an endurance function $W$, Eq. (\ref{eq:W}, which considers the way lifetime is distributed  to different components (parameter $\alpha$) and the functionality (parameter $\beta$) of the network during its lifetime. Our main finding is that for large $\beta$ (high connectivity) $\alpha_c$ -- the $\alpha$ value that maximizes $W$ -- tends to 0 and uniform lifetime division between the nodes is required. As $\beta$ decreases (require large survival time) $\alpha_c$ increases, i.e., there is a preference of lifetime allocation to high degree nodes. These findings could be useful for recognizing the actual users' requirements and correspondingly improving the network endurance in the design stage.\\

\acknowledgements{A.P., S.H. and R.C. acknowledge the Israel Science Foundation, Israel Ministry of Science and Technology (MOST) with the Italy Ministry of Foreign Affairs, MOST with the Japan Science and Technology Agency, ONR and DTRA for financial support. D.L. is supported by the National Natural Science Foundation of China (Grant No. 71771009). R.K. is supported by the National Natural Science Foundation of China (Grant No. 61573043). R.C. thanks the support of the BSF financial. Y.L. thanks the support from the program of China Scholarships Council (No. 201506020065).}

\bibliography{my_bib}

\begin{thebibliography}{42}%
\makeatletter
\providecommand \@ifxundefined [1]{%
 \@ifx{#1\undefined}
}%
\providecommand \@ifnum [1]{%
 \ifnum #1\expandafter \@firstoftwo
 \else \expandafter \@secondoftwo
 \fi
}%
\providecommand \@ifx [1]{%
 \ifx #1\expandafter \@firstoftwo
 \else \expandafter \@secondoftwo
 \fi
}%
\providecommand \natexlab [1]{#1}%
\providecommand \enquote  [1]{``#1''}%
\providecommand \bibnamefont  [1]{#1}%
\providecommand \bibfnamefont [1]{#1}%
\providecommand \citenamefont [1]{#1}%
\providecommand \href@noop [0]{\@secondoftwo}%
\providecommand \href [0]{\begingroup \@sanitize@url \@href}%
\providecommand \@href[1]{\@@startlink{#1}\@@href}%
\providecommand \@@href[1]{\endgroup#1\@@endlink}%
\providecommand \@sanitize@url [0]{\catcode `\\12\catcode `\$12\catcode
  `\&12\catcode `\#12\catcode `\^12\catcode `\_12\catcode `\%12\relax}%
\providecommand \@@startlink[1]{}%
\providecommand \@@endlink[0]{}%
\providecommand \url  [0]{\begingroup\@sanitize@url \@url }%
\providecommand \@url [1]{\endgroup\@href {#1}{\urlprefix }}%
\providecommand \urlprefix  [0]{URL }%
\providecommand \Eprint [0]{\href }%
\providecommand \doibase [0]{http://dx.doi.org/}%
\providecommand \selectlanguage [0]{\@gobble}%
\providecommand \bibinfo  [0]{\@secondoftwo}%
\providecommand \bibfield  [0]{\@secondoftwo}%
\providecommand \translation [1]{[#1]}%
\providecommand \BibitemOpen [0]{}%
\providecommand \bibitemStop [0]{}%
\providecommand \bibitemNoStop [0]{.\EOS\space}%
\providecommand \EOS [0]{\spacefactor3000\relax}%
\providecommand \BibitemShut  [1]{\csname bibitem#1\endcsname}%
\let\auto@bib@innerbib\@empty
\bibitem [{\citenamefont {Boccaletti}\ \emph {et~al.}(2006)\citenamefont
  {Boccaletti}, \citenamefont {Latora}, \citenamefont {Moreno}, \citenamefont
  {Chavez},\ and\ \citenamefont {Hwang}}]{boccaletti-complex-2006}%
  \BibitemOpen
  \bibfield  {author} {\bibinfo {author} {\bibfnamefont {S.}~\bibnamefont
  {Boccaletti}}, \bibinfo {author} {\bibfnamefont {V.}~\bibnamefont {Latora}},
  \bibinfo {author} {\bibfnamefont {Y.}~\bibnamefont {Moreno}}, \bibinfo
  {author} {\bibfnamefont {M.}~\bibnamefont {Chavez}}, \ and\ \bibinfo {author}
  {\bibfnamefont {D.-U.}\ \bibnamefont {Hwang}},\ }\href {\doibase
  10.1016/j.physrep.2005.10.009} {\bibfield  {journal} {\bibinfo  {journal}
  {Physics reports}\ }\textbf {\bibinfo {volume} {424}},\ \bibinfo {pages}
  {175} (\bibinfo {year} {2006})}\BibitemShut {NoStop}%
\bibitem [{\citenamefont {Dorogovtsev}\ \emph {et~al.}(2008)\citenamefont
  {Dorogovtsev}, \citenamefont {Goltsev},\ and\ \citenamefont
  {Mendes}}]{Dorogovtsev-RevModPhys-2008}%
  \BibitemOpen
  \bibfield  {author} {\bibinfo {author} {\bibfnamefont {S.~N.}\ \bibnamefont
  {Dorogovtsev}}, \bibinfo {author} {\bibfnamefont {A.~V.}\ \bibnamefont
  {Goltsev}}, \ and\ \bibinfo {author} {\bibfnamefont {J.~F.}\ \bibnamefont
  {Mendes}},\ }\href {\doibase 10.1103/RevModPhys.80.1275} {\bibfield
  {journal} {\bibinfo  {journal} {Reviews of Modern Physics}\ }\textbf
  {\bibinfo {volume} {80}},\ \bibinfo {pages} {1275} (\bibinfo {year}
  {2008})}\BibitemShut {NoStop}%
\bibitem [{\citenamefont {Strogatz}(2001)}]{strogatz-nature-2001}%
  \BibitemOpen
  \bibfield  {author} {\bibinfo {author} {\bibfnamefont {S.~H.}\ \bibnamefont
  {Strogatz}},\ }\href {\doibase 10.1038/35065725} {\bibfield  {journal}
  {\bibinfo  {journal} {nature}\ }\textbf {\bibinfo {volume} {410}},\ \bibinfo
  {pages} {268} (\bibinfo {year} {2001})}\BibitemShut {NoStop}%
\bibitem [{\citenamefont {Albert}\ and\ \citenamefont
  {Barab{\'a}si}(2002)}]{albert-RevModPhys-2002}%
  \BibitemOpen
  \bibfield  {author} {\bibinfo {author} {\bibfnamefont {R.}~\bibnamefont
  {Albert}}\ and\ \bibinfo {author} {\bibfnamefont {A.}~\bibnamefont
  {Barab{\'a}si}},\ }\href {\doibase 10.1103/RevModPhys.74.47} {\bibfield
  {journal} {\bibinfo  {journal} {Rev. Mod. Phys.}\ }\textbf {\bibinfo {volume}
  {74}},\ \bibinfo {pages} {47} (\bibinfo {year} {2002})}\BibitemShut {NoStop}%
\bibitem [{\citenamefont {Albert}\ \emph {et~al.}(2000)\citenamefont {Albert},
  \citenamefont {Jeong},\ and\ \citenamefont
  {Barab{\'a}si}}]{albert-nature-2000}%
  \BibitemOpen
  \bibfield  {author} {\bibinfo {author} {\bibfnamefont {R.}~\bibnamefont
  {Albert}}, \bibinfo {author} {\bibfnamefont {H.}~\bibnamefont {Jeong}}, \
  and\ \bibinfo {author} {\bibfnamefont {A.-L.}\ \bibnamefont {Barab{\'a}si}},\
  }\href {\doibase 10.1038/35019019} {\bibfield  {journal} {\bibinfo  {journal}
  {Nature}\ }\textbf {\bibinfo {volume} {406}},\ \bibinfo {pages} {378}
  (\bibinfo {year} {2000})}\BibitemShut {NoStop}%
\bibitem [{\citenamefont {Cohen}\ \emph {et~al.}(2000)\citenamefont {Cohen},
  \citenamefont {Erez}, \citenamefont {Ben-Avraham},\ and\ \citenamefont
  {Havlin}}]{cohen-prl-2000}%
  \BibitemOpen
  \bibfield  {author} {\bibinfo {author} {\bibfnamefont {R.}~\bibnamefont
  {Cohen}}, \bibinfo {author} {\bibfnamefont {K.}~\bibnamefont {Erez}},
  \bibinfo {author} {\bibfnamefont {D.}~\bibnamefont {Ben-Avraham}}, \ and\
  \bibinfo {author} {\bibfnamefont {S.}~\bibnamefont {Havlin}},\ }\href
  {\doibase 10.1103/PhysRevLett.85.4626} {\bibfield  {journal} {\bibinfo
  {journal} {Physical review letters}\ }\textbf {\bibinfo {volume} {85}},\
  \bibinfo {pages} {4626} (\bibinfo {year} {2000})}\BibitemShut {NoStop}%
\bibitem [{\citenamefont {Cohen}\ \emph {et~al.}(2001)\citenamefont {Cohen},
  \citenamefont {Erez}, \citenamefont {Ben-Avraham},\ and\ \citenamefont
  {Havlin}}]{cohen-prl-2001}%
  \BibitemOpen
  \bibfield  {author} {\bibinfo {author} {\bibfnamefont {R.}~\bibnamefont
  {Cohen}}, \bibinfo {author} {\bibfnamefont {K.}~\bibnamefont {Erez}},
  \bibinfo {author} {\bibfnamefont {D.}~\bibnamefont {Ben-Avraham}}, \ and\
  \bibinfo {author} {\bibfnamefont {S.}~\bibnamefont {Havlin}},\ }\href
  {\doibase 10.1103/PhysRevLett.86.3682} {\bibfield  {journal} {\bibinfo
  {journal} {Physical review letters}\ }\textbf {\bibinfo {volume} {86}},\
  \bibinfo {pages} {3682} (\bibinfo {year} {2001})}\BibitemShut {NoStop}%
\bibitem [{\citenamefont {Zio}(2016)}]{zio-RelEng-2016}%
  \BibitemOpen
  \bibfield  {author} {\bibinfo {author} {\bibfnamefont {E.}~\bibnamefont
  {Zio}},\ }\href {\doibase 10.1016/j.ress.2016.02.009} {\bibfield  {journal}
  {\bibinfo  {journal} {Reliability Engineering \& System Safety}\ }\textbf
  {\bibinfo {volume} {152}},\ \bibinfo {pages} {137} (\bibinfo {year}
  {2016})}\BibitemShut {NoStop}%
\bibitem [{\citenamefont {Buldyrev}\ \emph {et~al.}(2010)\citenamefont
  {Buldyrev}, \citenamefont {Parshani}, \citenamefont {Paul}, \citenamefont
  {Stanley},\ and\ \citenamefont {Havlin}}]{buldyrev-nature-2010}%
  \BibitemOpen
  \bibfield  {author} {\bibinfo {author} {\bibfnamefont {S.~V.}\ \bibnamefont
  {Buldyrev}}, \bibinfo {author} {\bibfnamefont {R.}~\bibnamefont {Parshani}},
  \bibinfo {author} {\bibfnamefont {G.}~\bibnamefont {Paul}}, \bibinfo {author}
  {\bibfnamefont {H.~E.}\ \bibnamefont {Stanley}}, \ and\ \bibinfo {author}
  {\bibfnamefont {S.}~\bibnamefont {Havlin}},\ }\href {\doibase
  10.1038/nature08932} {\bibfield  {journal} {\bibinfo  {journal} {Nature}\
  }\textbf {\bibinfo {volume} {464}},\ \bibinfo {pages} {1025} (\bibinfo {year}
  {2010})}\BibitemShut {NoStop}%
\bibitem [{\citenamefont {Daqing}\ \emph {et~al.}(2014)\citenamefont {Daqing},
  \citenamefont {Yinan}, \citenamefont {Rui},\ and\ \citenamefont
  {Havlin}}]{daqing-scirep-2014l}%
  \BibitemOpen
  \bibfield  {author} {\bibinfo {author} {\bibfnamefont {L.}~\bibnamefont
  {Daqing}}, \bibinfo {author} {\bibfnamefont {J.}~\bibnamefont {Yinan}},
  \bibinfo {author} {\bibfnamefont {K.}~\bibnamefont {Rui}}, \ and\ \bibinfo
  {author} {\bibfnamefont {S.}~\bibnamefont {Havlin}},\ }\href {\doibase
  10.1038/srep05381} {\bibfield  {journal} {\bibinfo  {journal} {Scientific
  reports}\ }\textbf {\bibinfo {volume} {4}} (\bibinfo {year} {2014}),\
  10.1038/srep05381}\BibitemShut {NoStop}%
\bibitem [{\citenamefont {Dodds}\ \emph {et~al.}(2003)\citenamefont {Dodds},
  \citenamefont {Watts},\ and\ \citenamefont
  {Sabel}}]{Dodds-proc-nat-acad-sci-2003}%
  \BibitemOpen
  \bibfield  {author} {\bibinfo {author} {\bibfnamefont {P.~S.}\ \bibnamefont
  {Dodds}}, \bibinfo {author} {\bibfnamefont {D.~J.}\ \bibnamefont {Watts}}, \
  and\ \bibinfo {author} {\bibfnamefont {C.~F.}\ \bibnamefont {Sabel}},\ }\href
  {\doibase 10.1073/pnas.1534702100} {\bibfield  {journal} {\bibinfo  {journal}
  {Proceedings of the National Academy of Sciences}\ }\textbf {\bibinfo
  {volume} {100}},\ \bibinfo {pages} {12516} (\bibinfo {year}
  {2003})}\BibitemShut {NoStop}%
\bibitem [{\citenamefont {Li}\ \emph {et~al.}(2015)\citenamefont {Li},
  \citenamefont {Fu}, \citenamefont {Wang}, \citenamefont {Lu}, \citenamefont
  {Berezin}, \citenamefont {Stanley},\ and\ \citenamefont
  {Havlin}}]{li-proc-nat-acad-sci-2015}%
  \BibitemOpen
  \bibfield  {author} {\bibinfo {author} {\bibfnamefont {D.}~\bibnamefont
  {Li}}, \bibinfo {author} {\bibfnamefont {B.}~\bibnamefont {Fu}}, \bibinfo
  {author} {\bibfnamefont {Y.}~\bibnamefont {Wang}}, \bibinfo {author}
  {\bibfnamefont {G.}~\bibnamefont {Lu}}, \bibinfo {author} {\bibfnamefont
  {Y.}~\bibnamefont {Berezin}}, \bibinfo {author} {\bibfnamefont {H.~E.}\
  \bibnamefont {Stanley}}, \ and\ \bibinfo {author} {\bibfnamefont
  {S.}~\bibnamefont {Havlin}},\ }\href {\doibase 10.1073/pnas.1419185112}
  {\bibfield  {journal} {\bibinfo  {journal} {Proceedings of the National
  Academy of Sciences}\ }\textbf {\bibinfo {volume} {112}},\ \bibinfo {pages}
  {669} (\bibinfo {year} {2015})}\BibitemShut {NoStop}%
\bibitem [{\citenamefont {Chowdhury}\ \emph {et~al.}(2000)\citenamefont
  {Chowdhury}, \citenamefont {Santen},\ and\ \citenamefont
  {Schadschneider}}]{CHOWDHURY-phys-rep-2000}%
  \BibitemOpen
  \bibfield  {author} {\bibinfo {author} {\bibfnamefont {D.}~\bibnamefont
  {Chowdhury}}, \bibinfo {author} {\bibfnamefont {L.}~\bibnamefont {Santen}}, \
  and\ \bibinfo {author} {\bibfnamefont {A.}~\bibnamefont {Schadschneider}},\
  }\href {\doibase 10.1016/S0370-1573(99)00117-9} {\bibfield  {journal}
  {\bibinfo  {journal} {Physics Reports}\ }\textbf {\bibinfo {volume} {329}},\
  \bibinfo {pages} {199} (\bibinfo {year} {2000})}\BibitemShut {NoStop}%
\bibitem [{\citenamefont {Kitano}(2004)}]{kitano-nature-2004}%
  \BibitemOpen
  \bibfield  {author} {\bibinfo {author} {\bibfnamefont {H.}~\bibnamefont
  {Kitano}},\ }\href {\doibase 10.1038/nrg1471} {\bibfield  {journal} {\bibinfo
   {journal} {Nature Reviews Genetics}\ }\textbf {\bibinfo {volume} {5}},\
  \bibinfo {pages} {826} (\bibinfo {year} {2004})}\BibitemShut {NoStop}%
\bibitem [{\citenamefont {Kitano}(2002)}]{kitano-science-2002}%
  \BibitemOpen
  \bibfield  {author} {\bibinfo {author} {\bibfnamefont {H.}~\bibnamefont
  {Kitano}},\ }\href {\doibase 10.1126/science.1069492} {\bibfield  {journal}
  {\bibinfo  {journal} {Science}\ }\textbf {\bibinfo {volume} {295}},\ \bibinfo
  {pages} {1662} (\bibinfo {year} {2002})}\BibitemShut {NoStop}%
\bibitem [{\citenamefont {Folke}(2006)}]{folke-glob-envir-chng-2006}%
  \BibitemOpen
  \bibfield  {author} {\bibinfo {author} {\bibfnamefont {C.}~\bibnamefont
  {Folke}},\ }\href {\doibase 10.1016/j.gloenvcha.2006.04.002} {\bibfield
  {journal} {\bibinfo  {journal} {Global environmental change}\ }\textbf
  {\bibinfo {volume} {16}},\ \bibinfo {pages} {253} (\bibinfo {year}
  {2006})}\BibitemShut {NoStop}%
\bibitem [{\citenamefont {Anderies}\ \emph {et~al.}(2004)\citenamefont
  {Anderies}, \citenamefont {Janssen},\ and\ \citenamefont
  {Ostrom}}]{anderies-ecology-society-2004}%
  \BibitemOpen
  \bibfield  {author} {\bibinfo {author} {\bibfnamefont {J.}~\bibnamefont
  {Anderies}}, \bibinfo {author} {\bibfnamefont {M.}~\bibnamefont {Janssen}}, \
  and\ \bibinfo {author} {\bibfnamefont {E.}~\bibnamefont {Ostrom}},\
  }\href@noop {} {\bibfield  {journal} {\bibinfo  {journal} {Ecology and
  society}\ }\textbf {\bibinfo {volume} {9}} (\bibinfo {year}
  {2004})}\BibitemShut {NoStop}%
\bibitem [{\citenamefont {Newman}\ \emph {et~al.}(2002)\citenamefont {Newman},
  \citenamefont {Watts},\ and\ \citenamefont
  {Strogatz}}]{newman-nat-acad-sci-2002}%
  \BibitemOpen
  \bibfield  {author} {\bibinfo {author} {\bibfnamefont {M.~E.}\ \bibnamefont
  {Newman}}, \bibinfo {author} {\bibfnamefont {D.~J.}\ \bibnamefont {Watts}}, \
  and\ \bibinfo {author} {\bibfnamefont {S.~H.}\ \bibnamefont {Strogatz}},\
  }\href {\doibase 10.1073/pnas.012582999} {\bibfield  {journal} {\bibinfo
  {journal} {Proceedings of the National Academy of Sciences}\ }\textbf
  {\bibinfo {volume} {99}},\ \bibinfo {pages} {2566} (\bibinfo {year}
  {2002})}\BibitemShut {NoStop}%
\bibitem [{\citenamefont {Eubank}\ \emph {et~al.}(2004)\citenamefont {Eubank},
  \citenamefont {Guclu}, \citenamefont {Kumar}, \citenamefont {Marathe} \emph
  {et~al.}}]{eubank-nature-2004}%
  \BibitemOpen
  \bibfield  {author} {\bibinfo {author} {\bibfnamefont {S.}~\bibnamefont
  {Eubank}}, \bibinfo {author} {\bibfnamefont {H.}~\bibnamefont {Guclu}},
  \bibinfo {author} {\bibfnamefont {V.~A.}\ \bibnamefont {Kumar}}, \bibinfo
  {author} {\bibfnamefont {M.~V.}\ \bibnamefont {Marathe}},  \emph {et~al.},\
  }\href {\doibase 10.1038/nature02541} {\bibfield  {journal} {\bibinfo
  {journal} {Nature}\ }\textbf {\bibinfo {volume} {429}},\ \bibinfo {pages}
  {180} (\bibinfo {year} {2004})}\BibitemShut {NoStop}%
\bibitem [{\citenamefont {Li}(2002)}]{li-trans-power-sys-2002}%
  \BibitemOpen
  \bibfield  {author} {\bibinfo {author} {\bibfnamefont {W.}~\bibnamefont
  {Li}},\ }\href {\doibase 10.1109/TPWRS.2002.800989} {\bibfield  {journal}
  {\bibinfo  {journal} {IEEE Transactions on Power systems}\ }\textbf {\bibinfo
  {volume} {17}},\ \bibinfo {pages} {918} (\bibinfo {year} {2002})}\BibitemShut
  {NoStop}%
\bibitem [{\citenamefont {Ko{\l}owrocki}\ and\ \citenamefont
  {Kwiatuszewska-Sarnecka}(2008)}]{KOLOWROCKI-rel-eng-sys-2008}%
  \BibitemOpen
  \bibfield  {author} {\bibinfo {author} {\bibfnamefont {K.}~\bibnamefont
  {Ko{\l}owrocki}}\ and\ \bibinfo {author} {\bibfnamefont {B.}~\bibnamefont
  {Kwiatuszewska-Sarnecka}},\ }\href {\doibase 10.1016/j.ress.2008.03.008}
  {\bibfield  {journal} {\bibinfo  {journal} {Reliability Engineering \& System
  Safety}\ }\textbf {\bibinfo {volume} {93}},\ \bibinfo {pages} {1821}
  (\bibinfo {year} {2008})}\BibitemShut {NoStop}%
\bibitem [{\citenamefont {Finkel}\ and\ \citenamefont
  {Holbrook}(2000)}]{finkel-nature-2000}%
  \BibitemOpen
  \bibfield  {author} {\bibinfo {author} {\bibfnamefont {T.}~\bibnamefont
  {Finkel}}\ and\ \bibinfo {author} {\bibfnamefont {N.~J.}\ \bibnamefont
  {Holbrook}},\ }\href {\doibase 10.1038/35041687} {\bibfield  {journal}
  {\bibinfo  {journal} {Nature}\ }\textbf {\bibinfo {volume} {408}},\ \bibinfo
  {pages} {239} (\bibinfo {year} {2000})}\BibitemShut {NoStop}%
\bibitem [{\citenamefont {Harman}(1981)}]{Harman-proc-nat-acad-sci-1981}%
  \BibitemOpen
  \bibfield  {author} {\bibinfo {author} {\bibfnamefont {D.}~\bibnamefont
  {Harman}},\ }\href@noop {} {\bibfield  {journal} {\bibinfo  {journal}
  {Proceedings of the National Academy of Sciences}\ }\textbf {\bibinfo
  {volume} {78}},\ \bibinfo {pages} {7124} (\bibinfo {year}
  {1981})}\BibitemShut {NoStop}%
\bibitem [{\citenamefont {Viswanath}\ \emph {et~al.}(2009)\citenamefont
  {Viswanath}, \citenamefont {Mislove}, \citenamefont {Cha},\ and\
  \citenamefont {Gummadi}}]{viswanath-2009}%
  \BibitemOpen
  \bibfield  {author} {\bibinfo {author} {\bibfnamefont {B.}~\bibnamefont
  {Viswanath}}, \bibinfo {author} {\bibfnamefont {A.}~\bibnamefont {Mislove}},
  \bibinfo {author} {\bibfnamefont {M.}~\bibnamefont {Cha}}, \ and\ \bibinfo
  {author} {\bibfnamefont {K.~P.}\ \bibnamefont {Gummadi}},\ }in\ \href
  {\doibase 10.1145/1592665.1592675} {\emph {\bibinfo {booktitle} {Proceedings
  of the 2nd ACM workshop on Online social networks}}}\ (\bibinfo
  {organization} {ACM},\ \bibinfo {year} {2009})\ pp.\ \bibinfo {pages}
  {37--42}\BibitemShut {NoStop}%
\bibitem [{\citenamefont {Leskovec}\ \emph {et~al.}(2008)\citenamefont
  {Leskovec}, \citenamefont {Backstrom}, \citenamefont {Kumar},\ and\
  \citenamefont {Tomkins}}]{Leskovec-2008}%
  \BibitemOpen
  \bibfield  {author} {\bibinfo {author} {\bibfnamefont {J.}~\bibnamefont
  {Leskovec}}, \bibinfo {author} {\bibfnamefont {L.}~\bibnamefont {Backstrom}},
  \bibinfo {author} {\bibfnamefont {R.}~\bibnamefont {Kumar}}, \ and\ \bibinfo
  {author} {\bibfnamefont {A.}~\bibnamefont {Tomkins}},\ }in\ \href {\doibase
  10.1145/1401890.1401948} {\emph {\bibinfo {booktitle} {Proceedings of the
  14th ACM SIGKDD international conference on Knowledge discovery and data
  mining}}}\ (\bibinfo {organization} {ACM},\ \bibinfo {year} {2008})\ pp.\
  \bibinfo {pages} {462--470}\BibitemShut {NoStop}%
\bibitem [{\citenamefont {Berezin}\ \emph {et~al.}(2015)\citenamefont
  {Berezin}, \citenamefont {Bashan}, \citenamefont {Danziger}, \citenamefont
  {Li},\ and\ \citenamefont {Havlin}}]{berezin-sci-rep-2015}%
  \BibitemOpen
  \bibfield  {author} {\bibinfo {author} {\bibfnamefont {Y.}~\bibnamefont
  {Berezin}}, \bibinfo {author} {\bibfnamefont {A.}~\bibnamefont {Bashan}},
  \bibinfo {author} {\bibfnamefont {M.~M.}\ \bibnamefont {Danziger}}, \bibinfo
  {author} {\bibfnamefont {D.}~\bibnamefont {Li}}, \ and\ \bibinfo {author}
  {\bibfnamefont {S.}~\bibnamefont {Havlin}},\ }\href {\doibase
  10.1038/srep08934} {\bibfield  {journal} {\bibinfo  {journal} {Scientific
  reports}\ }\textbf {\bibinfo {volume} {5}} (\bibinfo {year} {2015}),\
  10.1038/srep08934}\BibitemShut {NoStop}%
\bibitem [{\citenamefont {Paul}\ \emph {et~al.}(2004)\citenamefont {Paul},
  \citenamefont {Tanizawa}, \citenamefont {Havlin},\ and\ \citenamefont
  {Stanley}}]{paul-europ-phys-journal-2004}%
  \BibitemOpen
  \bibfield  {author} {\bibinfo {author} {\bibfnamefont {G.}~\bibnamefont
  {Paul}}, \bibinfo {author} {\bibfnamefont {T.}~\bibnamefont {Tanizawa}},
  \bibinfo {author} {\bibfnamefont {S.}~\bibnamefont {Havlin}}, \ and\ \bibinfo
  {author} {\bibfnamefont {H.~E.}\ \bibnamefont {Stanley}},\ }\href {\doibase
  10.1140/epjb/e2004-00112-3} {\bibfield  {journal} {\bibinfo  {journal} {The
  European Physical Journal B-Condensed Matter and Complex Systems}\ }\textbf
  {\bibinfo {volume} {38}},\ \bibinfo {pages} {187} (\bibinfo {year}
  {2004})}\BibitemShut {NoStop}%
\bibitem [{\citenamefont {Shargel}\ \emph {et~al.}(2003)\citenamefont
  {Shargel}, \citenamefont {Sayama}, \citenamefont {Epstein},\ and\
  \citenamefont {Bar-Yam}}]{shargel-prl-2003}%
  \BibitemOpen
  \bibfield  {author} {\bibinfo {author} {\bibfnamefont {B.}~\bibnamefont
  {Shargel}}, \bibinfo {author} {\bibfnamefont {H.}~\bibnamefont {Sayama}},
  \bibinfo {author} {\bibfnamefont {I.~R.}\ \bibnamefont {Epstein}}, \ and\
  \bibinfo {author} {\bibfnamefont {Y.}~\bibnamefont {Bar-Yam}},\ }\href
  {\doibase 10.1103/PhysRevLett.90.068701} {\bibfield  {journal} {\bibinfo
  {journal} {Physical review letters}\ }\textbf {\bibinfo {volume} {90}},\
  \bibinfo {pages} {068701} (\bibinfo {year} {2003})}\BibitemShut {NoStop}%
\bibitem [{\citenamefont {Tanizawa}\ \emph {et~al.}(2005)\citenamefont
  {Tanizawa}, \citenamefont {Paul}, \citenamefont {Cohen}, \citenamefont
  {Havlin},\ and\ \citenamefont {Stanley}}]{tanizawa-pre-2005}%
  \BibitemOpen
  \bibfield  {author} {\bibinfo {author} {\bibfnamefont {T.}~\bibnamefont
  {Tanizawa}}, \bibinfo {author} {\bibfnamefont {G.}~\bibnamefont {Paul}},
  \bibinfo {author} {\bibfnamefont {R.}~\bibnamefont {Cohen}}, \bibinfo
  {author} {\bibfnamefont {S.}~\bibnamefont {Havlin}}, \ and\ \bibinfo {author}
  {\bibfnamefont {H.~E.}\ \bibnamefont {Stanley}},\ }\href {\doibase
  10.1103/PhysRevE.71.047101} {\bibfield  {journal} {\bibinfo  {journal}
  {Physical review E}\ }\textbf {\bibinfo {volume} {71}},\ \bibinfo {pages}
  {047101} (\bibinfo {year} {2005})}\BibitemShut {NoStop}%
\bibitem [{\citenamefont {Braunewell}\ and\ \citenamefont
  {Bornholdt}(2008)}]{Braunewell-pre-2008}%
  \BibitemOpen
  \bibfield  {author} {\bibinfo {author} {\bibfnamefont {S.}~\bibnamefont
  {Braunewell}}\ and\ \bibinfo {author} {\bibfnamefont {S.}~\bibnamefont
  {Bornholdt}},\ }\href {\doibase 10.1103/PhysRevE.77.060902} {\bibfield
  {journal} {\bibinfo  {journal} {Physical Review E}\ }\textbf {\bibinfo
  {volume} {77}},\ \bibinfo {pages} {060902} (\bibinfo {year}
  {2008})}\BibitemShut {NoStop}%
\bibitem [{\citenamefont {Klemm}\ and\ \citenamefont
  {Bornholdt}(2005)}]{Klemm-proc-nat-acad-sci-2005}%
  \BibitemOpen
  \bibfield  {author} {\bibinfo {author} {\bibfnamefont {K.}~\bibnamefont
  {Klemm}}\ and\ \bibinfo {author} {\bibfnamefont {S.}~\bibnamefont
  {Bornholdt}},\ }\href {\doibase 10.1073/pnas.0509132102} {\bibfield
  {journal} {\bibinfo  {journal} {Proceedings of the National Academy of
  Sciences of the United States of America}\ }\textbf {\bibinfo {volume}
  {102}},\ \bibinfo {pages} {18414} (\bibinfo {year} {2005})}\BibitemShut
  {NoStop}%
\bibitem [{\citenamefont {Kauffman}\ and\ \citenamefont
  {Smith}(1986)}]{kauffman-physicaD-1986}%
  \BibitemOpen
  \bibfield  {author} {\bibinfo {author} {\bibfnamefont {S.~A.}\ \bibnamefont
  {Kauffman}}\ and\ \bibinfo {author} {\bibfnamefont {R.~G.}\ \bibnamefont
  {Smith}},\ }\href {\doibase 10.1016/0167-2789(86)90234-4} {\bibfield
  {journal} {\bibinfo  {journal} {Physica D: Nonlinear Phenomena}\ }\textbf
  {\bibinfo {volume} {22}},\ \bibinfo {pages} {68} (\bibinfo {year}
  {1986})}\BibitemShut {NoStop}%
\bibitem [{\citenamefont {Aldana}\ and\ \citenamefont
  {Cluzel}(2003)}]{Aldana-proc-nat-acad-sci-2003}%
  \BibitemOpen
  \bibfield  {author} {\bibinfo {author} {\bibfnamefont {M.}~\bibnamefont
  {Aldana}}\ and\ \bibinfo {author} {\bibfnamefont {P.}~\bibnamefont
  {Cluzel}},\ }\href {\doibase 10.1073/pnas.1536783100} {\bibfield  {journal}
  {\bibinfo  {journal} {Proceedings of the National Academy of Sciences}\
  }\textbf {\bibinfo {volume} {100}},\ \bibinfo {pages} {8710} (\bibinfo {year}
  {2003})}\BibitemShut {NoStop}%
\bibitem [{\citenamefont {Schneider}\ \emph {et~al.}(2011)\citenamefont
  {Schneider}, \citenamefont {Moreira}, \citenamefont {Andrade}, \citenamefont
  {Havlin},\ and\ \citenamefont {Herrmann}}]{Schneider-proc-nat-acad-sci-2011}%
  \BibitemOpen
  \bibfield  {author} {\bibinfo {author} {\bibfnamefont {C.~M.}\ \bibnamefont
  {Schneider}}, \bibinfo {author} {\bibfnamefont {A.~A.}\ \bibnamefont
  {Moreira}}, \bibinfo {author} {\bibfnamefont {J.~S.}\ \bibnamefont
  {Andrade}}, \bibinfo {author} {\bibfnamefont {S.}~\bibnamefont {Havlin}}, \
  and\ \bibinfo {author} {\bibfnamefont {H.~J.}\ \bibnamefont {Herrmann}},\
  }\href {\doibase 10.1073/pnas.1009440108} {\bibfield  {journal} {\bibinfo
  {journal} {Proceedings of the National Academy of Sciences}\ }\textbf
  {\bibinfo {volume} {108}},\ \bibinfo {pages} {3838} (\bibinfo {year}
  {2011})}\BibitemShut {NoStop}%
\bibitem [{\citenamefont {Saroiu}\ \emph {et~al.}(2002)\citenamefont {Saroiu},
  \citenamefont {Gummadi}, \citenamefont {Gribble} \emph
  {et~al.}}]{saroiu-2002}%
  \BibitemOpen
  \bibfield  {author} {\bibinfo {author} {\bibfnamefont {S.}~\bibnamefont
  {Saroiu}}, \bibinfo {author} {\bibfnamefont {P.~K.}\ \bibnamefont {Gummadi}},
  \bibinfo {author} {\bibfnamefont {S.~D.}\ \bibnamefont {Gribble}},  \emph
  {et~al.},\ }in\ \href@noop {} {\emph {\bibinfo {booktitle} {proceedings of
  Multimedia Computing and Networking}}},\ Vol.\ \bibinfo {volume} {2002}\
  (\bibinfo {year} {2002})\ p.\ \bibinfo {pages} {152}\BibitemShut {NoStop}%
\bibitem [{\citenamefont {Leonard}\ \emph {et~al.}(2005)\citenamefont
  {Leonard}, \citenamefont {Rai},\ and\ \citenamefont
  {Loguinov}}]{Leonard-perf-eval-rev-2005}%
  \BibitemOpen
  \bibfield  {author} {\bibinfo {author} {\bibfnamefont {D.}~\bibnamefont
  {Leonard}}, \bibinfo {author} {\bibfnamefont {V.}~\bibnamefont {Rai}}, \ and\
  \bibinfo {author} {\bibfnamefont {D.}~\bibnamefont {Loguinov}},\ }\href
  {\doibase 10.1145/1071690.1064217} {\bibfield  {journal} {\bibinfo  {journal}
  {ACM SIGMETRICS performance evaluation review}\ }\textbf {\bibinfo {volume}
  {33}},\ \bibinfo {pages} {26} (\bibinfo {year} {2005})}\BibitemShut {NoStop}%
\bibitem [{\citenamefont {Proschan}(1963)}]{porschan-technometrics-1963}%
  \BibitemOpen
  \bibfield  {author} {\bibinfo {author} {\bibfnamefont {F.}~\bibnamefont
  {Proschan}},\ }\href@noop {} {\bibfield  {journal} {\bibinfo  {journal}
  {Technometrics}\ }\textbf {\bibinfo {volume} {5}},\ \bibinfo {pages} {375}
  (\bibinfo {year} {1963})}\BibitemShut {NoStop}%
\bibitem [{\citenamefont {Pecht}(2008)}]{pecht-2008}%
  \BibitemOpen
  \bibfield  {author} {\bibinfo {author} {\bibfnamefont {M.}~\bibnamefont
  {Pecht}},\ }\href {\doibase 10.1002/9780470061626.shm118} {\emph {\bibinfo
  {title} {Prognostics and health management of electronics}}}\ (\bibinfo
  {publisher} {Wiley Online Library},\ \bibinfo {year} {2008})\BibitemShut
  {NoStop}%
\bibitem [{\citenamefont {Zio}(2007)}]{zio-2007}%
  \BibitemOpen
  \bibfield  {author} {\bibinfo {author} {\bibfnamefont {E.}~\bibnamefont
  {Zio}},\ }\href@noop {} {\emph {\bibinfo {title} {An introduction to the
  basics of reliability and risk analysis}}},\ Vol.~\bibinfo {volume} {13}\
  (\bibinfo  {publisher} {World scientific},\ \bibinfo {year}
  {2007})\BibitemShut {NoStop}%
\bibitem [{\citenamefont {Weibull}(1951)}]{weibull-jrnl-appl-mech-1951}%
  \BibitemOpen
  \bibfield  {author} {\bibinfo {author} {\bibfnamefont {W.}~\bibnamefont
  {Weibull}},\ }\href@noop {} {\bibfield  {journal} {\bibinfo  {journal}
  {Journal of applied mechanics}\ }\textbf {\bibinfo {volume} {103}},\ \bibinfo
  {pages} {293} (\bibinfo {year} {1951})}\BibitemShut {NoStop}%
\bibitem [{\citenamefont {Callaway}\ \emph {et~al.}(2000)\citenamefont
  {Callaway}, \citenamefont {Newman}, \citenamefont {Strogatz},\ and\
  \citenamefont {Watts}}]{callaway-prl-2000}%
  \BibitemOpen
  \bibfield  {author} {\bibinfo {author} {\bibfnamefont {D.~S.}\ \bibnamefont
  {Callaway}}, \bibinfo {author} {\bibfnamefont {M.~E.}\ \bibnamefont
  {Newman}}, \bibinfo {author} {\bibfnamefont {S.~H.}\ \bibnamefont
  {Strogatz}}, \ and\ \bibinfo {author} {\bibfnamefont {D.~J.}\ \bibnamefont
  {Watts}},\ }\href {\doibase 10.1103/PhysRevLett.85.5468} {\bibfield
  {journal} {\bibinfo  {journal} {Physical review letters}\ }\textbf {\bibinfo
  {volume} {85}},\ \bibinfo {pages} {5468} (\bibinfo {year}
  {2000})}\BibitemShut {NoStop}%
\bibitem [{\citenamefont {Morone}\ and\ \citenamefont
  {Makse}(2015)}]{morone-nature-2015}%
  \BibitemOpen
  \bibfield  {author} {\bibinfo {author} {\bibfnamefont {F.}~\bibnamefont
  {Morone}}\ and\ \bibinfo {author} {\bibfnamefont {H.~A.}\ \bibnamefont
  {Makse}},\ }\href {\doibase 10.1038/nature14604} {\bibfield  {journal}
  {\bibinfo  {journal} {Nature}\ }\textbf {\bibinfo {volume} {524}},\ \bibinfo
  {pages} {65} (\bibinfo {year} {2015})}\BibitemShut {NoStop}%
\end{thebibliography}%

\clearpage

\end{document}